\mag=\magstephalf
\pageno=1
\input amstex
\documentstyle{amsppt}
\TagsOnRight

\pagewidth{16.5 truecm}
\pageheight{23.0 truecm}
\vcorrection{-1.0cm}
\hcorrection{-0.5cm}
\nologo
\pageno=0
\NoRunningHeads
\NoBlackBoxes
\font\twobf=cmbx12

\define \ee{\roman e}

\define \tr{\roman {tr}}

\define \al{\roman {al}}

\define \CC{{\Bbb C}}
\define \RR{{\Bbb R}}

\define \diag{\roman {diag}}

\define \SD{\roman {SD}}

\define \Sym{{\roman{Sym}}}
\define\hw{\widehat{\omega}}

\define \tvskip{\vskip 1.0 cm}
\define\ce#1{\lceil#1\rceil}
\define\dg#1{(d^{\circ}\geq#1)}
\define\Dg#1#2{(d^{\circ}(#1)\geq#2)}

\def\fp{\flushpar}

\define\s#1{\sigma_{#1}}
\define\tp#1{\negthinspace\left.\ ^t#1\right.}
\define\mrm#1{\text{\rm#1}}
\define\lr#1{^{\sssize\left(#1\right)}}

\redefine\qed{\hbox{\vrule height6pt width3pt depth0pt}}
\font\Large=cmr10 scaled \magstep5

{\centerline{\bf{Hyperelliptic Loop Solitons with  Genus $g$:   }}}

{\centerline{\bf{Investigations of a Quantized Elastica  }}}

\author
Shigeki MATSUTANI
\endauthor
\affil
8-21-1 Higashi-Linkan Sagamihara 228-0811 Japan
\endaffil \endtopmatter

\footnotetext{e-mail:RXB01142\@nifty.ne.jp}

\document

\centerline{\twobf Abstract }\tvskip

In the previous work (J. Geom. Phys.
{\bf{39}} (2001) 50-61), the closed loop solitons in a plane,
{\it i.e.}, loops whose curvatures obey the modified Korteweg-de Vries
equations,
were investigated for the case related to algebraic curves with
genera one and two.
This article is a generalization of the previous article to
those of hyperelliptic curves with general genera.
It was proved that the tangential angle of loop soliton
is expressed by the Weierstrass  hyperelliptic al function
for a given hyperelliptic curve $y^2 = f(x)$
with genus $g$.


{\centerline{\bf{2000 MSC: 35Q53, 53A04, 14H45, 14H70 }}}

{\centerline{\bf{PACS: 02.30.Gp,
05.20.Dd,
46.70.Hg,
05.45.Yv
 }}}


\vskip 1 cm
{\centerline{\bf{\S 1. Introduction}}}
\vskip 0.5 cm

\vskip 1.0 cm

This article is on loop solitons related to hyperelliptic curves
with higher genera as an extension of the previous report [Ma5].

In [Ma2], I proposed a problem of a quantized elastica
(ideal thin elastic curve), or
statistical mechanics of elasticas, which is a model of
a large polymer in a plane, such as DNA at finite temperature.
When a position of the elastica in a complex plane
$\CC$ is denoted by $Z: S^1 \hookrightarrow \CC$,
 the partition function of elastica is given by
$$
        \Cal Z[\beta] := \int D Z \exp( -\beta E[Z]), \tag 1-1
$$
where $\beta$ is the inverse of temperature,
$DZ$ is a certain functional measure and
$E[Z]$ is the Euler-Bernoulli functional energy,
$$
          E[Z] := \oint ds k(s)^2, \tag 1-2
$$
for the arclength $s$, {\it i.e.}, induced metric of the curve,
and its curvature $k$.

As shown in [Ma2-4], the partition function is completely
determined by the orbits of the modified Korteweg-de Vries
(MKdV) hierarchical equations.
As a curve obeying the MKdV equation is known as a
loop soliton due to [KIW, I], the quantized elastica
problem is a realization of the loop solitons.
It is worth while noting that even though the soliton theory
is studied in field of physics, there are not so many examples
that  soliton equation is connected with a physical model
 including its multi-soliton solutions.

In this article, we will consider hyperelliptic solutions
of the loop solitons or excited states of a quantized
elastica. Theorem 3-2 is our main theorem of this article.
There we give explicit solutions of  closed loop solitons in a plane
related to hyperelliptic curves with general genera.

We will base on the result of [Ma7]; there we show
hyperelliptic solutions of the MKdV equation
in terms of the theories of the hyperelliptic
functions which were developed in nineteenth
century [Ba1-3, Kl] and are recently re-evaluated [BEL1,2, Ma5, 6].
Following the idea mentioned in the discussion in [Ma5],
we extend the results in [Ma5] to the case of general genus
in terms of  Weierstrass's hyperelliptic al functions [Ba2 p.34, W].

As the elastica problem has a deep history [L, T1, 2], I believe
that one of its reasons is its naturalness.
In the derivations of the solutions, it turns out that
the elastica problem is very natural even from
a mathematical viewpoint. In Remark 3-3,
we will give comments on its naturalness.

Further as I mentioned in [Ma1], the elastica problem is closely
related to automorphic function theory even though
the solutions are constructed in Abelian variety.
In fact the Euler-Bernoulli energy functional
can be expressed by
the Schwarz derivative,
$$
          E[Z] = \oint ds \{Z,s\}_{\SD}. \tag 1-3
$$
In \S 4, we comment on its relation to automorphic
functions.

\vskip 0.5 cm

{\centerline{\bf{\S 2. Differentials of a Hyperelliptic Curve}}}

\vskip 0.5 cm

In this section, we will review the hyperelliptic  functions
following [\^O, BEL2, Ba1-3]  without explanations and proofs.

We denote the set of complex number by $\Bbb C$ and
the set of integers by $\Bbb Z$.

\proclaim{\fp Convention 2-1}\it
We deal with a hyperelliptic curve $X_g$  of genus $g$
$(g>0)$ given by the affine equation,
$$ \split
   y^2 &= f(x) \\  &= \lambda_{2g+1} x^{2g+1} +
\lambda_{2g} x^{2g}+\cdots  +\lambda_2 x^2
+\lambda_1 x+\lambda_0  \\
     &=P(x)Q(x), \\
\endsplit  \tag 2-1
$$
where
$$
\split
        f(x) &= (x-b_1)(x-b_2)\cdots(x-b_{2g})(x-b_{2g+1}),\\
        Q(x) &= (x-c_1)(x-c_2)\cdots(x-c_g)(x-c),\\
        P(x) &= (x-a_1)(x-a_2)\cdots(x-a_g), \\
\endsplit \tag 2-2
$$
$\lambda_{2g+1}\equiv1$, and $\lambda_j$'s,
$a_j$'s, $b_j$'s, $c_j$'s  and $c$ are  complex values.
\endproclaim

\proclaim{\fp Definition 2-2 [Ba1,2, BEL1,2 \^O]}\it

 \roster
For a point $(x_i, y_i)\in X_g$, we define the following
quantities.

\item  Let us denote the homology of a hyperelliptic
curve $X_g $ by,
$$
\roman{H}_1(X_g, \Bbb Z)
  =\bigoplus_{j=1}^g\Bbb Z\alpha_{j}
   \oplus\bigoplus_{j=1}^g\Bbb Z\beta_{j},
 \tag 2-3
$$
where these intersections are given as
$[\alpha_i, \alpha_j]=0$, $[\beta_i, \beta_j]=0$ and
$[\alpha_i, \beta_j]=\delta_{i,j}$.

\item The unnormalized differentials of the first kind are
defined by,
$$   d u^{(i)}_1 := \frac{ d x_i}{2y}, \quad
      d u^{(i)}_2 :=  \frac{x_i d x_i}{2y}, \quad \cdots, \quad
     d u^{(i)}_g :=\frac{x_i^{g-1} d x_i}{2 y}.
      \tag 2-4
$$

\item The unnormalized period matrices are defined by,
$$    \pmb{\omega}':=\left[\left(
      \int_{\alpha_{j}}d u^{(a)}_{i}\right)_{ij}\right],
\quad
      \pmb{\omega}'':=\left[
       \left(\int_{\beta_{j}}d u^{(a)}_{i}\right)_{ij}\right],
 \quad
    \pmb{\omega}:=\left[\matrix \pmb{\omega}' \\ \pmb{\omega}''
     \endmatrix\right].
  \tag 2-5
$$

\item The normalized period matrices are given by,
$$    \ ^t\left[\matrix \hw_{1}  \cdots & \hw_{g}
        \endmatrix\right]
       :={\pmb{\omega}'}^{-1}  \ ^t\left[\matrix
          d u^{(i)}_{1} & \cdots
   d u^{(i)}_{g}\endmatrix\right] ,\quad
   \pmb \tau:={\pmb{\omega}'}^{-1}\pmb{\omega}'',
   \quad
    \hat{\pmb{\omega}}:=\left[\matrix 1_g \\ \pmb \tau
     \endmatrix\right].
       \tag 2-6
$$

\item The unnormalized differentials of the second kind are
 defined by,
$$   d \tilde u^{(i)}_1 := \frac{x_i^g d x_i}{2y_i}, \quad
      d\tilde  u^{(i)}_2 :=  \frac{x_i^{g+1} d x_i}{2y_i},
       \quad \cdots,
 \quad
     d\tilde  u^{(i)}_g :=\frac{x_i^{2g-1} d x_i}{2 y_i},
      \tag 2-7
$$
and $d \bold r^{(i)} := (d r_1^{(i)}, d r_2^{(i)},
\cdots, d r_g^{(i)})$,
$$
     (d \bold r^{(i)}):=\Lambda \pmatrix d \bold u^{(i)}
           \\ d \tilde{\bold u}^{(i)}
 \endpmatrix,
     \tag 2-8
$$
where $\Lambda$ is $2g \times g$ matrix defined by
$$
\split
        \Lambda& =
\left(\matrix 0 & \lambda_3 & 2 \lambda_4 &
            3 \lambda_5 & \cdots &
          (g-1)\lambda_{g+1}& g \lambda_{g+2
          }& (g+1)\lambda_{g+3}\\
        \  & 0 &  \lambda_5 & 2\lambda_6 & \cdots &
            (g-2)\lambda_{g+2}& (g-1) \lambda_{g+3}&  g
            \lambda_{g+4}\\
    \ & \ &  0         &  \lambda_7 & \cdots &
            (g-3)\lambda_{g+3}& (g-2) \lambda_{g+4}
               & (g-1)\lambda_{g+5}\\
   \ & \ &  \        & \ & \ddots &
            \vdots & \vdots& \vdots\\
 \ &  & \text{\Large 0}        &  \ & \ &
            \lambda_{2g-2} &2 \lambda_{2g-1}& 3\lambda_{2g+1}\\
 \ &  & \        &  \ & \ &
          0  & \lambda_{2g+1}& 0  \endmatrix\right.\\
&\qquad \qquad
\left.\matrix \cdots & (2g-3)\lambda_{2g-1} & (2g-2)\lambda_{2g}
             & (2g-1) \lambda_{2g+1}\\
          \cdots & (2g-4)\lambda_{2g} & (2g-3)\lambda_{2g+1}
             & 0                    \\
          \cdots & (2g-5)\lambda_{2g+1} &        0
             &                     \\
\cdots & 0 &
             &                  \ \\
     \  & \ &
             &                     \\
\ & \ &  \text{\Large 0}
             &                  \ \\
\ & \ &  \
             &                  \ \endmatrix\right).
\endsplit \tag 2-9
$$

\item The complete hyperelliptic integral matrices
of the second kind are defined by,
$$    \pmb{\eta}':=\left[\left(
    \int_{\alpha_{j}}d r^{(a)}_{i}\right)_{ij}\right],
\quad
      \pmb{\eta}'':=\left[\left(
       \int_{\beta_{j}}d r^{(a)}_{i}\right)_{ij}\right],
 \quad
    \pmb{\omega}:=\left[\matrix \pmb{\omega}' \\ \pmb{\omega}''
     \endmatrix\right].
  \tag 2-10
$$
\item
By defining the Abel map for $g$-th symmetric product
of the curve $X_g$,
$$ \bold u:\roman{Sym}^g( X_g) \longrightarrow \Bbb C^g, \quad
      \left( u_k((Q_i)_{i=1,\cdots,g}):= \sum_{i=1}^g
       \int_\infty^{Q_i} d u^{(i)}_k, \quad k=1,\cdots,g \right),
      \tag 2-11
$$
the Jacobi variety  $\Cal J_g$
are defined as complex torus,
$$
   {\Cal J_g} := \Bbb C^g /{ \pmb{\Lambda}} .
     \tag 2-12
$$
Here  ${ \pmb{\Lambda}}$  is a
lattice generated by ${\pmb{\omega}}$.

\endroster
\endproclaim

\tvskip

\vskip 0.5 cm
\proclaim {Definition 2-3  }

\it
\roster
The coordinate in $\Bbb C^g$ for
 points $\{Q_i\equiv(x_i,y_i)\ | \ i=1,\cdots,g\ \}$
of the curve $y^2 = f(x)$ is given by,
$$
  u_j :=\sum_{i=1}^g\int^{(x_i,y_i)}_\infty d u_j^{(i)} ,
\quad d u_j = \sum_{i=1}^g d u_j^{(i)} .
    \tag 2-13
$$

\item Using the coordinate $u_j$, $\sigma$ functions,
which is a holomorphic
function over $\Bbb C^g$, is defined by
[Ba2, p.336,p350, BEL1, Kl],
$$ \sigma(u)=\sigma(u;X_g):
  =\ \gamma\roman{exp}(-\dfrac{1}{2}\ ^t\ u
  \pmb{\eta}'{\pmb{\omega}'}^{-1}u)
  \vartheta\negthinspace
  \left[\matrix \delta'' \\ \delta' \endmatrix\right]
  ({\pmb{\omega}'}^{-1}u ;\pmb \tau),
     \tag 2-14
$$
where $\gamma$ is a certain constant factor,
$$\vartheta\negthinspace\left[\matrix a \\ b \endmatrix\right]
     (z; \pmb \tau)
    :=\sum_{n \in \Bbb Z^g} \exp \left[2\pi \sqrt{-1}\left\{
    \dfrac 12 \ ^t\negthinspace (n+a)\pmb \tau(n+a)
    + \ ^t\negthinspace (n+a)(z+b)\right\}\right],
     \tag 2-15
$$
for $g$-dimensional vectors $a$ and $b$,
 and
$$
 \delta' :=\ ^t\left[\matrix \dfrac {g}{2} & \dfrac{g-1}{2}
       & \cdots
      & \dfrac {1}{2}\endmatrix\right],
   \quad \delta'':=\ ^t\left[\matrix \dfrac{1}{2} & \cdots
& \dfrac{1}{2}
   \endmatrix\right].
     \tag 2-16
$$

\item
Hyperelliptic
$\wp$-function is  defined by [Ba1, Ba2, Kl],
$$   \wp_{i j}(u):=-\dfrac{\partial^2}{\partial
   u_i\partial u_j}
   \log \sigma(u) ,
         \tag 2-17
$$
and Hyperelliptic $\zeta_i$ function is  defined by,
$$
       \zeta_i(u) :=\dfrac{\partial}{\partial u_j}
   \log \sigma(u) . \tag 2-18
$$

\item
Weierstrass hyperelliptic
$\al_r$-function is  defined by [Ba2 p.340, W]
$$   \al_r(u):=\gamma'\sqrt{F(b_r)},
         \tag 2-19
$$
where $\gamma'$ is a certain constant,
$$
   \split
        F(x)&:= (x-x_1) \cdots (x-x_g)\\
            &=: \gamma_g x^g + \gamma_{g-1} x^{g-1}
            +\cdots + \gamma_{0}.
   \endsplit
          \tag 2-20
$$

\endroster

\endproclaim

\proclaim{\fp Proposition 2-4 }\it

\roster

\item $\wp_{g i}$ $(i=1,\cdots,g)$ is elementary
symmetric functions of $\{x_1,x_2,\cdots, x_g\}$
 [Ba1, 2, 3, BEL1], i.e.,
$$
        F(x) = x^g-\sum_{i=1}^g \wp_{g i} x^{i-1}.
         \tag 2-21
$$

\item $\zeta_j(u)$ is expressed by [BEL2 p.33-35]
$$
        -\zeta_i(u) =\sum_{k=1}^g \int_{\infty}^{x_i}
              d r_i -\frac{1}{2} \det A_{g-i},
           \tag 2-22
$$
where
$$
A_{n+1} :=
\pmatrix e_1   & -1      &  \    & \ & \              &  & \\
         2 e_2 & -e_1    &  1    & \ &\text{\Large 0} &  & \\
         \vdots&\vdots  &\vdots &\ddots &            &  & \\
    (n-2)e_{n-2}&-e_{n-3}&e_{n-4}&\cdots & \pm 1          &  & \\
    (n-1)e_{n-1}&-e_{n-2}&e_{n-3}&\cdots &\pm e_1        &\mp 1  & \\
     n   e_{n}  &-e_{n-1}&e_{n-2}&\cdots &\pm e_2        &\mp  e_1 &\pm 1\\
     (n+1)d_{n+1}  &-d_{n}&d_{n-1}&\cdots &\pm d_3        &\mp  d_2&\pm d_1
 \endpmatrix \tag 2-23
$$
and $e_i:= \wp_{g,g+1-i}$ and  $d_i:= \wp_{g,g, g+1-i}
:=\partial \wp_{g,g+1-i}/\partial u_g$.

\endroster

\endproclaim

As we show later, (2-22) is very important relation in our
quantized elastica, which was found by
Buchstaber, Enolskii and Leykin [BEL2]  (Appendix);
according to [BEL2] Baker in [Ba1] gave a wrong relation.
We note that $F(x)$ is a generator of the elementary
symmetric functions and the matrix $A_n$ (2-23) contains
the matrix of the Newton formula as
its minor matrix. In fact in the derivation of (2-23),
$A_n$ plays the role which connects the elementary and power sum symmetric
 functions.

\proclaim{\fp Definition 2-5}\it
\roster
\item A polynomial associated with $F(x)$ is introduced by
$$
\split
\pi_i(x) &:= \frac{F(x)}{x-x_i}\\
        &=\chi_{i,g-1}x^{g-1} +\chi_{i,g-2} x^{g-2}
            +\cdots+\chi_{i,1}+\chi_{i,0},\\
\endsplit \tag 2-24
$$
where $\chi_{i,g-1}\equiv1$, $\chi_{i,g-2}= (x^1+\cdots+x_g)-x_i$,
and so on.

\item We will introduce $g\times g$-matrices
$$
 W := \pmatrix
     \chi_{1,0} & \chi_{1,1} & \cdots & \chi_{1,g-1}  \\
      \chi_{2,0} & \chi_{2,1} & \cdots & \chi_{2,g-1}  \\
   \vdots & \vdots & \ddots & \vdots  \\
    \chi_{g,0} & \chi_{g,1} & \cdots & \chi_{g,g-1}
     \endpmatrix,\quad
        \Cal Y = \pmatrix
     y_1 & \ & \ & \  \\
      \ & y_2& \ & \   \\
      \ & \ & \ddots   & \   \\
      \ & \ & \ & y_g  \endpmatrix,
$$
$$
        \Cal F' = \pmatrix
     F'(x_1) & \ & \ & \  \\
      \ & F'(x_2)& \ & \   \\
      \ & \ & \ddots   & \   \\
      \ & \ & \ & F'(x_{g})  \endpmatrix,\quad
\tag 2-25
$$
where $F'(x):=d F(x)/d x$.

\item
$$
M:= \pmatrix
      1         & \ & \ & \ &\  \\
      \gamma_{g-1} & 1 & \ &\text{\Large 0}  & \  \\
   \gamma_{g-2} &\gamma_{g-1}     & 1      &  & \ \\
    \vdots& \vdots          & \vdots & \ddots & \\
   \gamma_{1} &\gamma_{2}     & \cdots      & \gamma_{g-1} & 1 \\
     \endpmatrix, \quad
K := \pmatrix
     x_1^{g-1} &   x_1^{g-2} & \cdots & 1  \\
  x_2^{g-1} &   x_2^{g-2} & \cdots & 1  \\
   \vdots & \vdots & \ddots & \vdots  \\
  x_g^{g-1} &   x_g^{g-2} & \cdots & 1
     \endpmatrix.\tag 2-26
$$

\item The coordinate in $\CC^g$ is introduced by
$\bold u^{(r)}:= \Cal P_r \bold u$, where $ \Cal P_r$
is defined by its inverse matrix,
$$
        \Cal P_{r}^{-1} = \pmatrix
     1 & g b_r & \pmatrix g-1\\2\endpmatrix b_r^2 & \cdots &
         \pmatrix g-1\\g-1\endpmatrix b_r^{g-1} & b_r^{g-1}   \\
     0 & 1 & (g-1) b_r & \cdots &
         \pmatrix g-1\\g-2\endpmatrix b_r^{g-3} & b_r^{g-2}   \\
      \vdots &\vdots & \vdots & \ddots   & \vdots & \vdots \\
     0 & 0 &0 & \cdots & 1 & b_r  \\
       0 & 0 &0 & \cdots & 0 & 1  \endpmatrix.\quad
\tag 2-27
$$

\item For a polynomial $g(X)= g_{n}X^n + \cdots + g_0$,
 we introduce the $D_j$ operator
$$
          D_j = \sum_{i=j}^n g_{i} X^{i-j}. \tag 2-28
$$

\endroster
\endproclaim

\proclaim{\fp Lemma 2-6}\it

\roster

\item The inverse matrix of $W$ is given by
$W^{-1}={\Cal F}^{-1} V$,
where $V$ is Vandermonde matrix,
$$
        V= \pmatrix 1 & 1 & \cdots & 1 \\
                   x_1 & x_2 & \cdots & x_g \\
                   x_1^2 & x_2^2 & \cdots & x_g^2 \\
                    \cdot& \cdot &       & \cdot \\
                   x_1^{g-1} & x_2^{g-1} & \cdots & x_g^{g-1}
                 \endpmatrix. \tag 2-29
$$

\item Let $\partial_{u_i}:=\partial/\partial{u_i}$,
$\partial_{x_i}:=\partial/\partial{x_i}$ and
$\partial_{u_i}^{(r)}:=\partial/\partial{u_i^{(r)}}$,

$$
        \pmatrix \partial_{u_1}\\
                 \partial_{u_2}\\
                 \vdots\\
                 \partial_{u_g}
         \endpmatrix
   =2 \Cal Y\Cal F^{\prime -1}W
        \pmatrix \partial_{x_1}\\
                 \partial_{x_2}\\
                 \vdots\\
                 \partial_{x_g}
         \endpmatrix, \quad
        \pmatrix \partial_{u_1}^{(r)}\\
                 \partial_{u_2}^{(r)}\\
                 \vdots\\
                 \partial_{u_g}^{(r)}
         \endpmatrix
   ={}^t\Cal P^{ -1}
\pmatrix \partial_{u_1}\\
                 \partial_{u_2}\\
                 \vdots\\
                 \partial_{u_g}
         \endpmatrix. \tag 2-30
$$

\item $KM = W$ and
$$
        W_{i j}= \chi_{i,j-1}= [D_j (F(X))]_{X=x_i}. \tag 2-31
$$
\endroster
\endproclaim

\demo{Proof}
(1) is obvious by using the properties of the Vandermonde matrix.
In (2), we must pay attention the fixed parameters for the partial
differential.  By comparing $dx^i$ and the chain relation of
$\partial_{u_i}$, we obtain the matrix representation (2-30) [Ma7].
From the relation $\left(\dfrac{F(x)}{x-x_i} \right)\cdot (x-x_i)
=F(x)$, we have
$$
        \chi_{i,j}= \gamma_{j+1} + x_i \chi_{i,j+1}.
\tag 2-32
$$
Then we obtain the relations (2-31).\qed \enddemo

We note that the formulae (2-31) and (2-32) are
 very important to prove
(2-23).

\proclaim{\fp Proposition 2-7 [BEL2 p.11 ]}\it

The Legendre relation is given by
$$
  {}^t\omega'\eta''- {}^t\omega''\eta' = 2\pi\sqrt{-1}I_g, \tag 2-33
$$
where $I_g$ is the $g\times g$-unit matrix.

\endproclaim

\tvskip
\centerline{\twobf \S 3. Loop Solitons }\tvskip

In this section, we will deal with a real curve in a plane
in the category of differential geometry.

Let us consider a smooth immersion of a circle $S^1$ into the two
dimensional Euclidean space $\Bbb E^2 \approx \Bbb C$ or
$\Bbb E^2 +\{\infty\}\approx \CC P^1$. The immersed
real curve $C$ is characterized by the affine coordinate
$(X^1(s),X^2(s))$ around the origin.
Here $s$ is a parameter of $S^1$ and is, now,
chosen as the arclength
so that $ds^2 = (dX^1)^2 + (d X^2)^2$.
We will also use the complex expression,
$$
        Z(s):= X^1(s)+ \sqrt{-1} X^2. \tag 3-1
$$
Then  by letting $\partial_s:=\partial/\partial s$,
$|\partial_s Z(s)|=1$ and the curvature of $C$ is given by
$$
      k(s):= \frac{1}{\sqrt{-1}} \partial_s \log \partial_s
      Z(s). \tag 3-2
$$
As mentioned in the introduction, loop soliton is identified with
a quantized elastica [Ma2]. Thus we will sometimes call it
{\it quantized elastica} or simply {\it elastica} hereafter.

\proclaim{\fp Definition 3-1}\it

\roster

\item
A one parameter family of curves $\{C_t\}$ for real parameter
$t\in \Bbb R$
is called a loop soliton,
if its curvature  obeys the MKdV equation;
for $q:=k/2$,
$$
        \partial_t q + 6 q^2 \partial_s q + \partial_s^3 q
         =0, \tag 3-3
$$
where $\partial_t :=\partial/\partial t$.

\item
The energy of elastica is given by
$$
        E[Z]:=\oint d s k^2, \tag 3-4
$$
which can be expressed by the Schwarz derivative,
$$
        E[Z]= \oint d s \{Z,s\}_{\SD}, \tag 3-5
$$
where
$$
        \{Z,s\}_{\SD}:=\partial_s\left(\frac{\partial_s^2 Z}
 {\partial_s Z}\right)
       -\frac{1}{2} \left(\frac{\partial_s^2 Z}
             {\partial_s Z}\right)^2
       .  \tag 3-6
$$

\endroster
\endproclaim

Here we will give our main theorem.

\proclaim{\fp Theorem 3-2}\it

\roster
Let the configuration of the $x$-components
$(x_1, \cdots, x_g)$ of the affine coordinates
of the hyperelliptic curves $\Sym^g(X_g)$
satisfy,
$$
        | F(b_r) | =r_0 ,\tag 3-7
$$
where $r_0$  is a positive number.
For such $(x_1,y_1),\cdots,(x_g,y_g)$, we have
$\bold u:=\bold u( (x_1,y_1),$ $\cdots,(x_g,y_g) )$
due to (2-11).

\item By setting $s\equiv  u_g/r_0$ and $t\equiv u_{g-1} +
( \lambda_{2g-1} +b_r) u_g$,
$$
   \partial_{u_g}Z^{(r)} :=  F(b_r), \quad
\text{or} \quad |\partial_s Z^{(r)} | = 1,  \tag 3-8
$$
completely characterizes the loop soliton.

\item The shape of elastica is given by,
$$
      Z^{(r)}= \frac{1}{r_0}\left(b_r^g u_g + \sum_{i=1}^g
                b_r^i \zeta_{i-1}\right). \tag 3-9
$$

\endroster

\endproclaim

\proclaim{\fp Remark 3-3}\rm

\roster

\item If one prefers more proper expression for
the branch point $(b_r,0)$, he may use
$u_j^{(r)}$ in (2-27) and then find similar results.
As the expression is essentially the same as the above
one due to (2-30), we will investigate the above one.

\item The condition (3-7) is essential. Due to the condition,
any configurations of $\bold u$, the tangential angle
$\phi=\log (\partial_{u_g} Z)/\sqrt{-1}$ does not
contain imaginary part. Hence the arclength locally does not
change [Ma1-4]. As Goldstein and Petrich showed that the isometric
deformation of space curve in a plane gives the MKdV equation
[GP],
this condition and $[\partial_{u_g}, \partial_{u_i}] =0$
recover the MKdV equation in general.

\item The tangential vector,
 $\partial_{u_g} Z \equiv F(b_r)\propto \roman{al}_r^2$,
consists only of $x_i$'s, which can be
regarded as a two-fold coordinate of $\Sym^g(\CC P^1)$.
Each $x_i\in \CC P^1$ appears when
 we construct the hyperelliptic curve $X_g$
using two $\CC P^1$ with $g+1$ cuts.

\item Due to the configuration of (3-7), there is a trivial
 action of the U(1)-group, which exhibits the translation
symmetry of the elastica. When we begin with this symmetry and
isometric deformation, we reproduces the MKdV hierarchy [Ma2,3].

\item
The condition (3-7) should be regarded as a reality condition.
Elastica problem is a real analytic problem. In the primitive
sense, the complex analysis is more complex than the real
analysis but from deeper viewpoint, their standpoints
are reversed. In fact, due to the condition, we must investigate
all possible contours in the complex curve $X_g$.
In other words, from the point of view of real analysis,
as long as we have insufficient knowledge
of the condition, it is not the end of the study
of the quantized elastica problem. I suppose that
this difficulty is similar to that of
real analytic Eisenstein series [S].

\item
The condition (3-7) is satisfied if all $x_i$ are in a
circle centrizing at $b_r$.
Then symmetric configuration of $x_1$, $\cdots$, $x_g$
determines a point of
a shape of the loop soliton. In other words,
the dynamics of the elastica is translated to symmetric system of
$g$-particles in $S^1$. Dynamics of symmetric
particles in a circle
might be familiar with researches of quantum
integrable system [KBI].

When we consider the discrete configurations of $x$'s,
they  give the discrete time development of the
piecewise linear curves. This must be related to
the discrete integrable system.

\endroster

\endproclaim

From the definition, Theorem 3-2 can be proved by
the next proposition, which was shown in [Ma7].
We will give a sketch of the proof of [Ma7],
whose techniques essentially appeared in [Ba3].

\proclaim{\fp Proposition 3-4}\it

By letting
$$\mu^{(r)} := \dfrac{1}{2} \partial_{u_g} \phi^{(r)},
 \quad
   \phi^{(r)}(u) :=\frac{1}{\sqrt{-1}} \log F(b_r),
        \tag 3-10
$$
$\mu^{(r)}$ obeys the modified KdV equation,
$$
        (\partial_{u_{g-1}}-(\lambda_{2g}+b_r)
          \partial_{u_{g}})\mu^{(r)}
           -6 {\mu^{(r)}}^2 \partial_{u_g} \mu^{(r)}
 +\partial_{u_g}^3 \mu^{(r)}=0.
      \tag 3-11
$$

\endproclaim

\demo{Proof}
This is proved by [Ma7]. We will give a sketch of the proof.
From the definition, we have
$$
        \frac{\partial}{\partial u_g }\log F(b_r)
         =\sum_{i=1}^g\frac{2y_i}{F'(x_i) (x_i-b_r)},
 \quad
        \frac{\partial}{\partial u_{g-1} }\log F(b_r)
         =\sum_{i=1}^g\frac{2y_i\chi_{i,g-1}}{F'(x_i)( x_i-b_r)}
          . \tag 3-12
$$
Let $\partial X_g^o$ is boundary when $X_g$ is embedded in
a upper half plane $\Cal H$.
By estimation of
$$
  \oint_{\partial X_g^o} \frac{f(x)}{(x-b_r)F(x)^2} dx =0, \tag 3-13
$$
and counting its residues, we obtain
$$
\sum_{k=1}^g \frac{1}{F'(x_k)}
             \left[\frac{\partial}{\partial x}\left(
       \frac{f(x)}{(x - b_r) F^2(x)} \right) \right]_{x = x_k}
       = \lambda_{2g} + b_r + 2 \wp_{gg}
         . \tag 3-14
$$
Further we have
$$
\left( \sum_k \frac{y_k}{(x-x_k) F'(x_k)} \right)^2
  =\sum_{k,l, k\neq l}
     \frac{2y_k y_l}{(x-x_k) (x_k - x_l)F'(x_k)F'(x_l)}
  +\sum_k \frac{y_k^2}{(x-x_k)^2 F'(x_k)^2}
         . \tag 3-15
$$
Using them we have the relation. \qed \enddemo

We note that the formal power sires
$$
        \mu^{(r)}\equiv \frac{1}{2\sqrt{-1}}
        \frac{\partial}{\partial u_g }\log F(b_r)
         =\frac{1}{\sqrt{-1}}\sum_{i=1}^g\frac{y_i}{F'(x_i) (x_i-b_r)}
        =\frac{1}{\sqrt{-1}}\sum_{j=1}^\infty
     \sum_{i=1}^g\frac{y_i}{F'(x_i) b_r }
\frac{ x_i ^j}{b_r^j}, \tag 3-16
$$
is resemble to the generator of the power sum symmetric functions.

\vskip 0.5 cm

As we proved Proposition 3-4, we give two corollaries,
which are shown by direct computations. Corollary 3-5
gives local properties of the elastica and Corollary 3-6
is associated with its global properties.

\proclaim{\fp Corollary 3-5}\it

\roster

\item The shape of elastica is given by

$$
      Z^{(r)}=\frac{1}{r_0}\left( b_r^g u_g + \sum_{i,j=1}^g
                b_r^i \int^{(x_j,y_j)} d r_{i-1}
       + \frac{1}{2}  \sum_{i,j=1}^g b_r^i \det A_{g-i}\right)
       .
\tag 3-17
$$

\item The Schwarz derivative of $Z$ with respect to $u_g$
$$
 \{Z^{(r)},u_g\}_{\SD}=4\wp_{gg}+2\lambda_{2g}+2b_r.
       \tag 3-18
$$

\item The root square of the tangential vector
$\sqrt{ \partial_{u_g} Z^{(r)}} $ $\equiv {\roman{al}}_r$
is a solution of the Dirac equation or Frenet-Serret equation [Ma1],
$$
\pmatrix \partial_{u_g} & \mu^{(r)} \\
          \mu^{(r)}       & - \partial_{u_g} \endpmatrix
\pmatrix \sqrt{ \partial_{u_g} Z^{(r)}} \\
\sqrt{ -\partial_{u_g} Z^{(r)}} \endpmatrix =0.
       \tag 3-19
$$
\endroster

\endproclaim

Here we comment on (3-18) and (3-19).
First we note that the solution of the Dirac equation
consists of al-functions due to (2-19) and (3-8).
Second it is noted that from the definitions (3-6) and (3-10),
(3-18) agrees with the Miura transformation,
$$
\mu^{(r)2} +\sqrt{-1} \partial_{u_g} \mu^{(r)}
=2\wp_{gg}+\lambda_{2g}+b_r,
       \tag 3-20
$$
because the left hand side consists of the solutions of
the MKdV equation (3-11) whereas the right hand side obeys
the KdV equation [BEL1, Ma6].
Further it is obvious that (3-19) has the same data as the
Miura transformation (3-20) by operating the Dirac operator
twice [Ma7]. On the other hand
(3-19) can be expressed by
$$
\pmatrix  -\partial_{u_g}    & 0 \\
          0  &  \partial_{u_g}  \endpmatrix
\pmatrix \sqrt{ \partial_{u_g} Z^{(r)}} \\
\sqrt{ -\partial_{u_g} Z^{(r)}} \endpmatrix =
\pmatrix 0  & \mu^{(r)} \\
          \mu^{(r)}       & 0 \endpmatrix
\pmatrix \sqrt{ \partial_{u_g} Z^{(r)}} \\
\sqrt{ -\partial_{u_g} Z^{(r)}} \endpmatrix.
       \tag 3-21
$$
Here we can recognize that the left hand side
is an operation in analytic category while right
hand side is an operation as an endmorphims in
a commutative algebra. This relation is
essential in study of $\Cal D$-module,
due to the statements in p.12-13 in [Bj].

\proclaim{\fp Corollary 3-6}\it

\roster

\item The winding number of elastica can be computed for a
given  path by the integration,
$$
    w : = \frac{1}{2\pi}  \oint \partial_{u_g} \phi(u_g) d u_g
         .
       \tag 3-22
$$

\item The closed condition of elastica,
$$
      \oint \partial_{u_g} Z^{(r)} d u_g \equiv 0,
       \tag 3-23
$$ consists of the conditions
$$
       {b^{(r)}}^g\omega_i +\sum_{i=1}^g
                {b^{(r)}}^i \eta_{i-1}=0, \tag 3-24
$$
using the hyperelliptic integral (2-5) and (2-6)

\item
$$
        \{Z^{(r)},u_g\}_{\SD}d u_g =-4d\zeta_g
                  +2 ( \lambda_{2g}+b_r)d u_g
             . \tag 3-25
$$

\item
$$
 \oint_{\beta_a}\{Z^{(r)},u_g\}_{\SD}d u_g =-4\eta_{ag}''
            +2(\lambda_{2g}+b_r)\omega_{ag}''
             . \tag 3-26
$$
$$
 \oint_{\alpha_a}\{Z^{(r)},u_g\}_{\SD}d u_g =-4\eta_{ag}'
             +2(\lambda_{2g}+b_r)\omega_{ag}'
             . \tag 3-27
$$

\endroster

\endproclaim

\vskip 0.5 cm

{\centerline{\bf{\S 4. Discussion}}}

\vskip 0.5 cm

As mentioned in [Ma1], our problem is resemble to
Poincar\'e, Klein and Schwarz theories of the
automorphic function over the half plane or the Poincar\'e disk.

First the resemblance is obtained through the conformal
field theory.
Since $S^1$ is homotopically equivalent with $\CC-\{0\}$,
most results of the conformal field theory are obtained
by investigation of dynamics of functions over $S^1$ in a
Riemann sphere $\CC P^1$ [Ka].
The situation in (3-7) is very resemble to that of the conformal
field theory
but is in the higher genus Riemannian surface.
However we can also
 perform the Fourier transformation or localization
around the ramified point $(b_r,0)$. Then it should be noted
that on the injective maps,
$$
        \partial_s Z^{(r)}: S^1 \to X_g, \tag 4-1
$$
 there are nontrivial actions of
the fundamental group (3-23).
By letting
$$
        L_n := \oint \left(2\wp_{gg}+\lambda_{2g} + b_r\right)
          \ee^{2\pi\sqrt{-1} n u_g/r_0} \frac{d u_g}{r_0}, \tag 4-2
$$
we have the relation of Virasoro algebra [Be,Ma1],
$$
   [L_n, L_m] = (n-m) L_{n+m} + \delta_{0, n+m} n(n^2-1).
\tag 4-3
$$

Here we remark that in the conformal field theory, the surface
obtained by means of the Vertex operator acting a Riemannian
sphere is called a Riemannian surface with genus $g$ but
as in showed in [Ma6], such a Riemannian surface,
at least of the case of the hyperelliptic curve, is very
far from our Riemannian surfaces. The Riemannian surface
obtained by the Vertex operator from the Riemannian sphere
is, in fact, topologically genus $g$ surface but is
very special (semistable) degenerate curve; the theory over
such a curve
should be regarded as a theory on Riemannian sphere [AE].
Recently some of physicists might regard that topological
aspect in field theory is the most important and
they deal only with degenerate curves. However, at least,
in low energy physics related to our lives, we need
finer topology. In fact, the shape of classical elastica,
which was studied by Euler as a classical field theory and
is determined by curvature,
leads us to very fruitful physics and mathematics.
Further in general, physical phenomenon are not in a framework
of complex analysis. Even though some objects in the category
of the complex analysis is classified by
topological objects, it is not all in physics.
For example, Euler equation for complete fluid dynamics
in three-dimensional space cannot be expressed by
complex analysis. Quantized elastica problem should be also
 considered in the real analytic category as Euler did.
Of course, topological aspect is still important but,
I believe, is not a goal for quantitative science.

Next I will comment on interesting relations, which also
looks connected with the automorphic functions.
$Z^{(r)}$ is roughly equal to $\zeta$-functions due to (3-17)
whereas the incomplete "energy integral",
$$
        \int^{\bold u}
        \{Z^{(r)},u_g\}_{\SD}d u_g, \tag 4-4
$$
is also expressed by the $\zeta$ functions from (3-18).
It implies that there is a similarity between the configuration
and energy of elasticas.
Further the matrix (2-23) is essentially the
same as the Newton formula which connects
the elementary and power sum symmetric functions.
Thus $\zeta$, (the configurations and energy from above view
points), is essentially expressed by the power sum symmetric
function, while the main part of $\mu$, or a half curvature of
 elastica,
 is the same as a generator of the power sum symmetric function
due to (3-16).
It is expected that there might be  hidden symmetries.
These facts might remind us of replicability of automorphic
function if they were related to automorphic functions [Mc].
In fact, for the case of elliptic function ($g=1$) case
helping with the notations in [Ma5],
($(\partial_u\wp(u))^2 = 4 \wp^3 -g_2 \wp + g_3$
$=4(\wp-e_1)(\wp- e_3)(\wp-e_4)$, $\wp(\omega_i)=e_i$ ),
 we have interesting
formulae,
$$
 \partial_{u} Z^{(a)}(u+\omega_a)
=\frac{1}{4} \{Z^{(a)}(u),u\}_{\SD}-\frac{3}{2}e_1, \tag 4-5
$$
$$
\split
        Z^{(a)}(u) =\lim_{\epsilon\to0} \int^u du
\frac{1}{\sigma(\epsilon)^2}\exp \Bigr(
-\frac{1}{2}\int^u_\epsilon \int^{u'}_0
\ & \Bigr[ \{Z^{(a)}(u''), u''\}_{\SD} \\
 &-\{Z^{(a)}(u''-\omega_a),u''\}_{\SD}\Bigr] du'' du' \Bigr),\\
\endsplit \tag 4-6
$$
for $a=1,2,3$. In the integrations, we should note the
effects from the initial points.
These equations (4-5) and (4-6)
can be extended to higher genus. For the case of (4-6)
we can do by using
the relation al$_r$ functions and $\sigma$ function [Ba3].

Further from the point of view of theory of
the symmetric function,
one might wish to regard  $x_i$ as an eigenvalue of some
matrix, $\Cal X = \diag(x_1, x_2,\cdots, x_g)$.
The generator of the elementary symmetric function
is expressed by,
$$
\tilde F(x) = \det( \Cal X - x I ), \tag 4-7
$$
and that of the power symmetric function is expressed by
$$
\tilde G(x) = \sum_{n=1}^\infty \tr (\Cal X^n) x^n .
\tag 4-8
$$
They are closely related to (2-20) and (3-16).
As comparison with physics, some purposes in mathematics are
classifications and determination of the relations among
classified objects.  The classification should be characterized by
discrete quantities and these discrete quantities should
sometimes preserve when we take a certain limit.
Thus it might be natural to consider degenerate curves in
a certain sense.
The degenerate curve
$y^2=P(x)^2 x$, which was dealt with in [Ma6]  and is
associated with the soliton solutions and algebra of vertex
operators, is also expressed by $y^2-x(\det( xI - \Cal P))^2=0$
by letting $\Cal P:=\diag(a_1,a_2,\cdots, a_g)$.
These matrices $\Cal P$ and $\Cal X$
for the degenerate curve are the same rank.
Hence it is natural that one consider a graded algebra $\Cal A$
($\Cal A_g \subset \Cal A_{g+1}$, $\Cal A = \cup_g \Cal A_g$)
generated by  generators ($\overline{\Cal X}$, $\overline{\Cal P}$
and so on) such that there is a $\Cal A$-module $M_g$ satisfying
$\overline{\Cal X}M_g = \Cal X M_g $ and
$\overline{\Cal P}M_g = \Cal P M_g$.
If we regard $M_g$ as a representation of shape of elastica,
 we might able to deal with family of elastica
related to hyperelliptic curves with different genera.
The Schwarz derivative, which plays important roles
in theory of automorphic functions [Mc] and is
 invariant for PSL$(2,\CC)$, naturally appears in our elastica
problem [Ma4].
Even though PSL$(2,\Bbb Z)$ is far from our situation in
this stage, I hope that
our study might reveal  some relations between elastica
problems and automorphic function theory [Mc], if exists [Ma1].

Finally we mention our future study. The quantized elastica
in a plane was extended to that in $\RR^3$. There
the nonlinear Schr\"odinger equation and complex MKdV
equation play the same role as the MKdV equation [Ma3].
As the explicit function form of the finite type solution
the nonlinear Schr\"odinger equation was obtained
in [EEK], this study might be extended to that in $\RR^3$.

\vskip 0.5 cm

{\centerline{\bf{ Acknowledgment}}}

\vskip 0.5 cm

I would like to grateful  Prof. Y.\^Onishi for
critical discussions and continuous encouragements.
I thank Prof. V. Z. Enolskii for sending me
their unpublished works, noticing the Newton formula,
explanations of (2-22) and helpful comments.
Further it should be acknowledgment that
 Prof. J. McKay is interested in this research
and pointed out the resemblance of this elastica
problem and his research related to  Monstrous Moonshine
and H. Mitsuhashi pointed out me the importance of
the Adams operator. Equation (4-6) was found
under the stimulus of a formula obtained by
Prof. J. McKay.

\vskip 0.5 cm

{\centerline{\bf{Appendix}}}

\vskip 0.5 cm

As (2-22) was given only in improved version [BEL2] of published
version [BEL1], we briefly review its derivation here following [BEL2].
There Buchstaber, Enolskii and Leykin started from a
fundamental  relation among the hyperelliptic sigma functions
and incomplete integrals of the first, second and third kinds.
Then they
used the operator (2-28), investigated of symmetry of an
intermediate equation and reached (2-22).

Thus let us follow their way.
First we will give the found-mental formula in the $\sigma$
function theory [Ba1, BEL1]
$$
\log\left(
\frac{\sigma\left(\int^{\mrm P}_\infty d \bold u + \bold u \right)
      \sigma\left(\int^{\mrm Q}_\infty d \bold u + \bold u'\right)}
     {\sigma\left(\int^{\mrm P}_\infty d \bold u + \bold u'\right)
      \sigma\left(\int^{\mrm Q}_\infty d \bold u + \bold u\right)}
      \right)
=\sum_{j=1}^g\bold{ R}
  _{\overline{\mrm P_j},
      \overline{\mrm Q_j}}^{\mrm P, \mrm Q},
       \tag A-1
  $$
where
$$
  \bold u=\sum_{j=1}^g\int_{\infty}^{\mrm P_j}d \bold u, \quad
  \bold u'=\sum_{j=1}^g\int_{\infty}^{\mrm Q_j}d \bold u,
   \tag A-2
$$
$$
\split
\bold{ R}_{{\mrm P'}, {\mrm Q'}}^{\mrm P, \mrm Q}
  &\equiv\bold{ R}_{{\mrm P}, {\mrm Q}}^{\mrm P', \mrm Q'}\\
  &=\int_{\mrm Q}^{\mrm P}d u_{1}\int_{\mrm Q'}^{\mrm P'}d r_{1}
   +\cdots
  +\int_{\mrm Q}^{\mrm P}d u_{g}\int_{\mrm Q'}^{\mrm P'}d r_{g}
  +\bold{P}_{\mrm Q, \mrm B}^{\mrm P, \mrm A} ,
 \endsplit  \tag A-3
$$
$$
  \bold{ P}_{\mrm P', \mrm Q'}^{\mrm P, \mrm Q}
:=\int_{\mrm Q}^{\mrm P}\Omega({\mrm P}',{\mrm Q}'),\quad
\Omega({\mrm P},{\mrm Q}):=\left(
          \frac{y+y_{\mrm P}}{x-x_{\mrm P} }
              -\frac{y+y_{\mrm Q}}{x-x_{\mrm Q} }\right)
             \frac{d x}{2y} ,
  \tag A-4
$$
and $\overline{\mrm P_j}$ $(\overline{\mrm Q_j})$
is conjugate of ${\mrm P_j}$ $({\mrm Q_j})$
with respect to the symmetry of hyperelliptic curve
$(x,y) \to (x,-y)$.

After letting all $\mrm Q_j$  to $\infty$,
we derivative it in $u_j$, we obtain
$$
\zeta_j( \int^{\mrm P}_\infty d \bold u + \bold u)
-\zeta_j( \int^{\mrm Q}_\infty d \bold u + \bold u)
 + \int^{\mrm P}_{\mrm Q} d r_j
- \frac{\partial}{\partial u_j}
\sum_{i=1}^g\int_\infty^{\overline{\mrm P}_i}
 \Omega(\mrm P, \mrm Q)=0.
\tag A-5
$$
Introducing $R(z)=(z-x_0)F(z)$, for $\mrm P=(x_0,y_0)$ and
$\mrm P_j=(x_j, y_j)$,
(A-5) becomes
$$
\split
&\zeta_j( \int^{\mrm P}_\infty d \bold u + \bold u)
 + \int^{\mrm P}_{\infty} d r_j+
\sum_{i=0}^g \int^{\mrm P_j}_{\infty} d r_j \\
&-\frac{1}{2}\sum_{i=0}^g
y_i\left(\frac{ D_j(R'(z))-jD_{j+1}(R(z))}
         {R'(z)} \Bigr|_{z=x_i} \right)\\
&=\zeta_j(\bold u)
 + \sum_{i=1}^g \int^{\mrm P_j}_{\infty} d r_j
-\frac{1}{2} \sum_{k=1}^g \frac{1}{y_k}
\frac{\partial x_k}{\partial u_j} \frac{1}{y_i}
           \frac{y_i - y_\infty}{x_i - x_\infty}\\
&-\frac{1}{2}\sum_{i=1}^g
y_i\left(\frac{ D_j(F'(z))-jD_{j+1}(F(z))}
         {R'(z)} \Bigr|_{z=x_i} \right).
\endsplit
\tag A-6
$$
Then noting the fact that
the left hand side is symmetrical in $x_0$, $x_1$, $\cdots$
$x_g$, while the right hand side is symmetrical in
$x_1$, $x_2$ $\cdots$ $x_g$ but does not depend on $x_0$,
(A-6) is reduced to (2-22). (2-23) comes from the third term
in the left hand side of (A-6).

\enddocument